\documentclass[11pt]{article}
\usepackage{moriond,epsfig}
\usepackage{subfig}

\def\Journal#1#2#3#4{{#1} {\bf #2}, #3 (#4)}

\def\PRB{{\em Phys. Rev.} B}

\def\be{\begin{equation}}
\def\ee{\end{equation}}
\def\bea{\begin{eqnarray}}
\def\eea{\end{eqnarray}}

\begin{document}
\vspace*{4cm}
\title{ANTARES: Status, first results and multi-messenger astronomy}

\author{Manuela Vecchi for the ANTARES Collaboration}

\address{
Universit\'{e} Aix-Marseille II and CPPM,\\
163, avenue de Luminy - Case 902 - 13288\\
Marseille cedex 09 \, France} 
\maketitle\abstracts{
The ANTARES Collaboration has completed in 2008 the deployment of what is currently the largest high energy neutrino detector in the Northern hemisphere, covering a volume of about 0.01 km$^3$.
The search for cosmic neutrinos in the energy range between tens of GeV and tens of PeV is 
performed by means of a three dimensional array of photomultiplier tubes (PMTs), arranged on 12 vertical structures (strings) 
located in the Mediterranean Sea at a depth of about 2500 meters. The 
detection principle relies on the 
identification of the Cherenkov light produced as ultra-relativistic muons propagate in water. 
The main goal of the detector is the search for point-like sources of cosmic neutrinos from both Galactic and extra-Galactic 
sources. 
Besides the search for point sources, other analysis topics are strongly pursued and will be described in the following.
}
\section{Introduction}
Cosmic rays (CRs) were discovered a century ago, but it is still uncertain where or how they are 
accelerated: multi-messenger 
astronomy~\cite{nu:review} could 
solve this \emph{puzzle}, combining the information coming from $\gamma$-rays, cosmic rays, neutrinos and gravitational 
waves. Charged particles, whose paths are deflected by magnetic fields, cannot carry the 
information on the arrival direction up to ultra high energies, so that a close look at production 
and acceleration sites of cosmic rays is only possible with neutral particles. 
Neutrinos are thought to be produced by the interaction of accelerated particles (protons 
and nuclei) with matter and radiation surrounding the sources. In these interactions a 
massive production of hadrons with short lifetime (mostly pions and kaons, both charged and 
neutral) is expected to take place, high energy neutrinos being their decay products.
Neutral hadrons, produced along with the charged particles generating neutrinos, are 
expected to decay into couples of high energy $\gamma$-rays, so that 
simultaneous emitters of neutrinos and $\gamma$-rays are very likely to exist. 
\section{Operation of the ANTARES Neutrino Telescope}
Cosmic neutrinos can be detected via the identification 
of the charged particles, in particular muons, that are produced as a consequence of charged current
interactions of neutrinos with the target matter. Relativistic muons propagating in a transparent medium, can 
induce the Cherenkov effect, i. e. the emission of coherent electromagnetic 
radiation along the surface of a cone, whose 
aperture is a function of the refraction index of the 
medium itself (about 42$^{\circ}$ for deep sea water). The detection technique relies on the observation
of Cherenkov radiation in the visible range, by means of a tridimensional array of photomultiplier tubes (PMTs).\\
The ANTARES Collaboration has completed the deployment of a neutrino telescope~\cite{antares_det} that is located about 
2500 meters deep, offshore Toulon, France.
The PMTs are arranged 
on 12 detection lines, each comprising up to 25 triplets of 
PMTs (floors), regularly distributed on 350~m, the lowest floor being located at 100 m above the sea bed.
Each line is connected to a junction box, which is itself connected to 
the shore station by a 40 km long electro-optical cable. The data collected on shore are then processed by a PC farm running 
several trigger algorithms looking for signals compatible with the ones produced 
by charged particles propagating through the detector. The counting rate of the detector, of the order of 100 kHz, is dominated by light emitted by bioluminescent bacteria and by the 
Cherenkov light that is emitted by electrons created as a decay product of radioactive elements present in sea water, such as $^{40}$K. 
Environmental background hits are mostly uncorrelated, and can be 
easily rejected by the trigger algorithm, which selects about 20 Hz of data.\\
The search for HE neutrinos is affected by a particle background, coming from the interactions of CRs with the upper
layers of the atmosphere, producing both neutrinos and muons, showing 
the same experimental signature of cosmic neutrinos.
\begin{figure}[!h]
\includegraphics[width=0.5\linewidth]{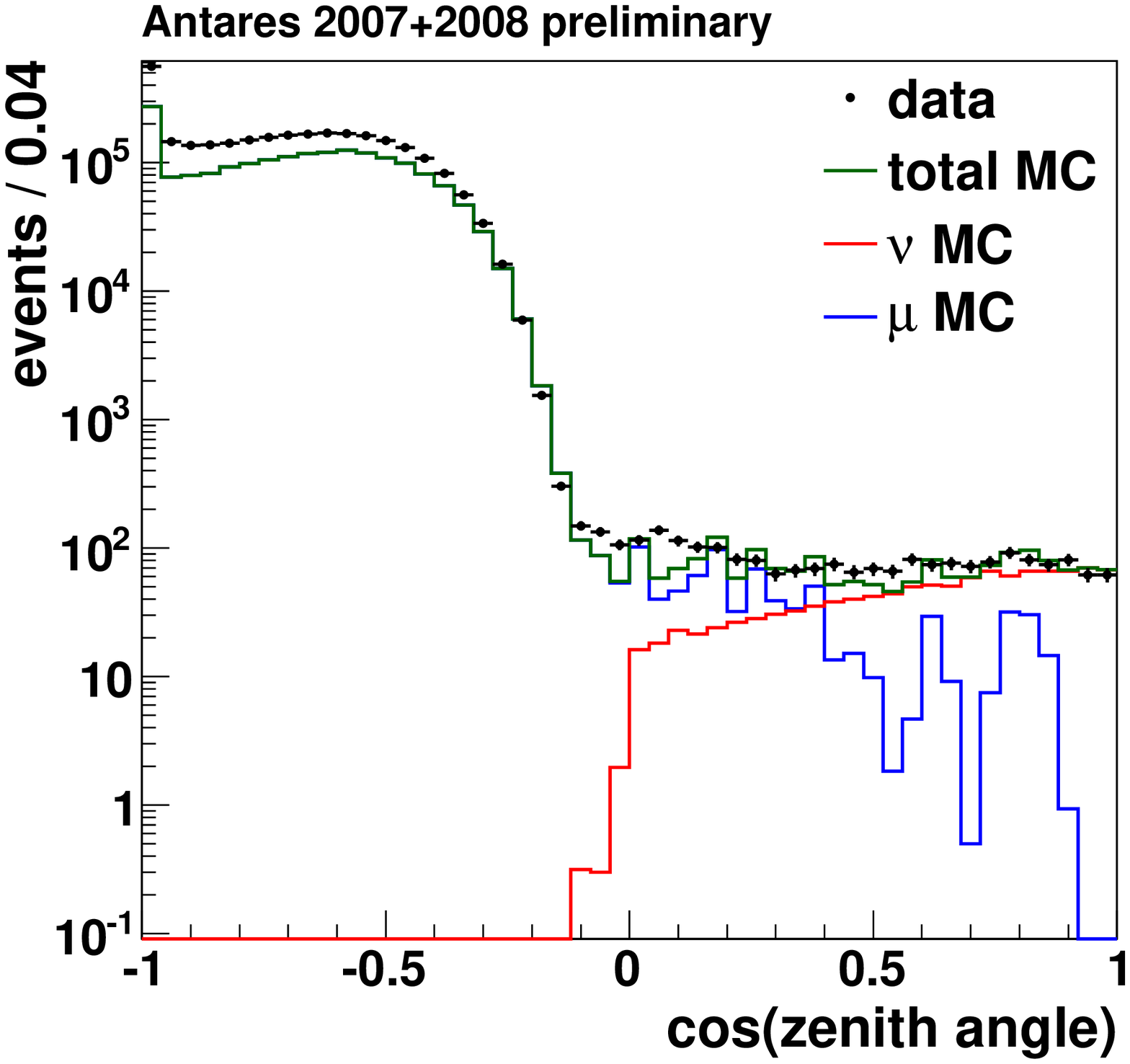}
\includegraphics[width=0.5\linewidth]{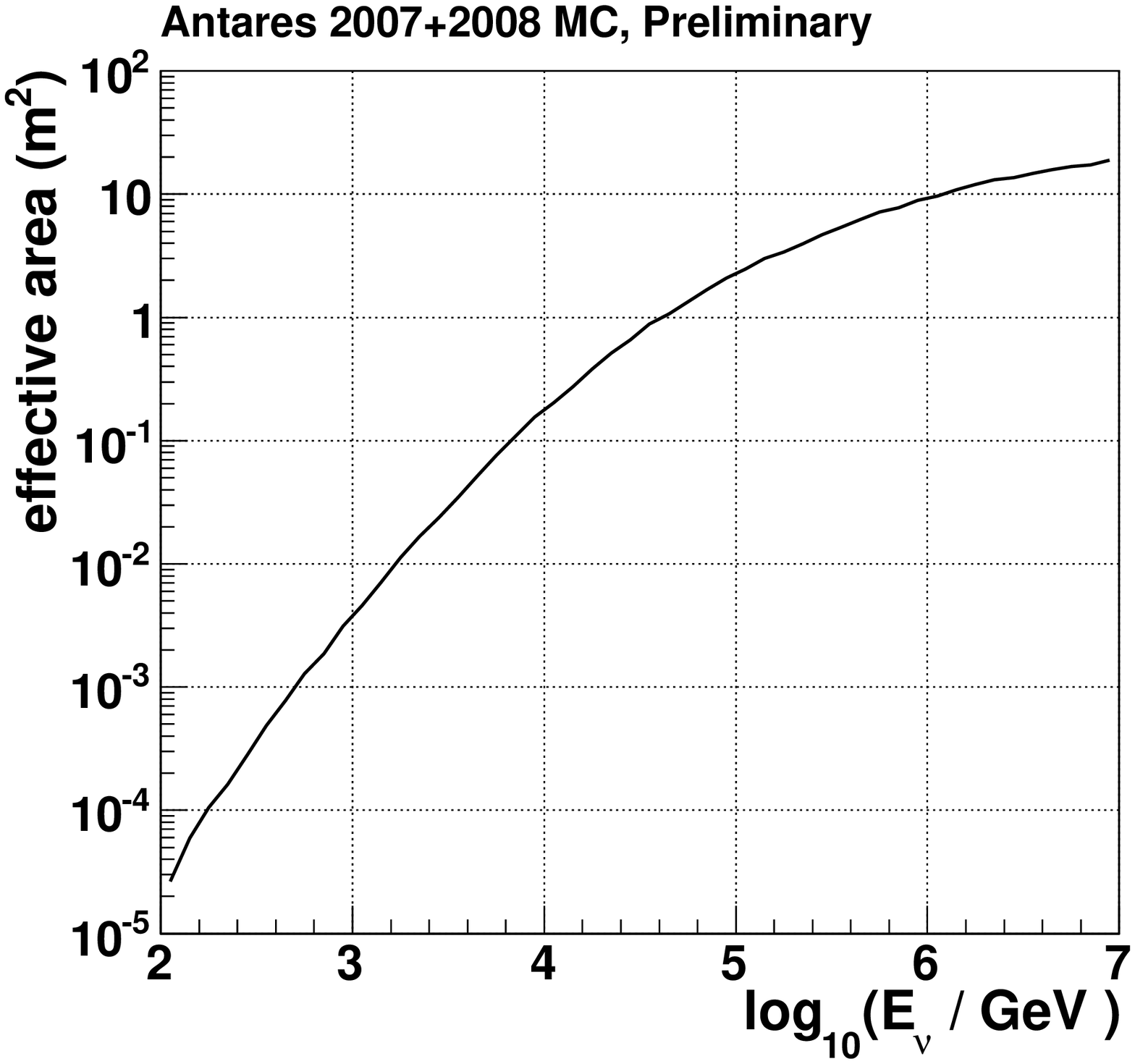}
\caption{\emph{Left}: Reconstructed arrival directions of the events detected with the ANTARES detector between 2007 and 2008. 
Events with cos$\theta <$ 0 are reconstructed as down-going, while events with cos$\theta >$ 0 are reconstructed as up-going.
\emph{Right}: Effective area of the ANTARES detector as a function of the neutrino energy, for an analysis 
optimized for point-like sources search.}
\label{2007-8:plot1}
\end{figure}    
Atmospheric muons, providing the most abundant flux, propagate downward through the detector, while
atmospheric neutrinos contribute 
providing an isotropic flux that is several orders of
magnitude less abundant, as can be seen in figure~\ref{2007-8:plot1}-\emph{Left}, showing 
the distribution of reconstructed arrival directions of the events detected with 
the ANTARES detector between 2007 and 2008, together with background expectations from simulations.
\section{Search for neutrinos from point-like sources}
\begin{figure}[!h]
\begin{minipage}[t]{0.5\textwidth}
\vspace{0pt}
\includegraphics[trim = 1mm 1mm 10mm 65mm, clip, width=0.9\linewidth]{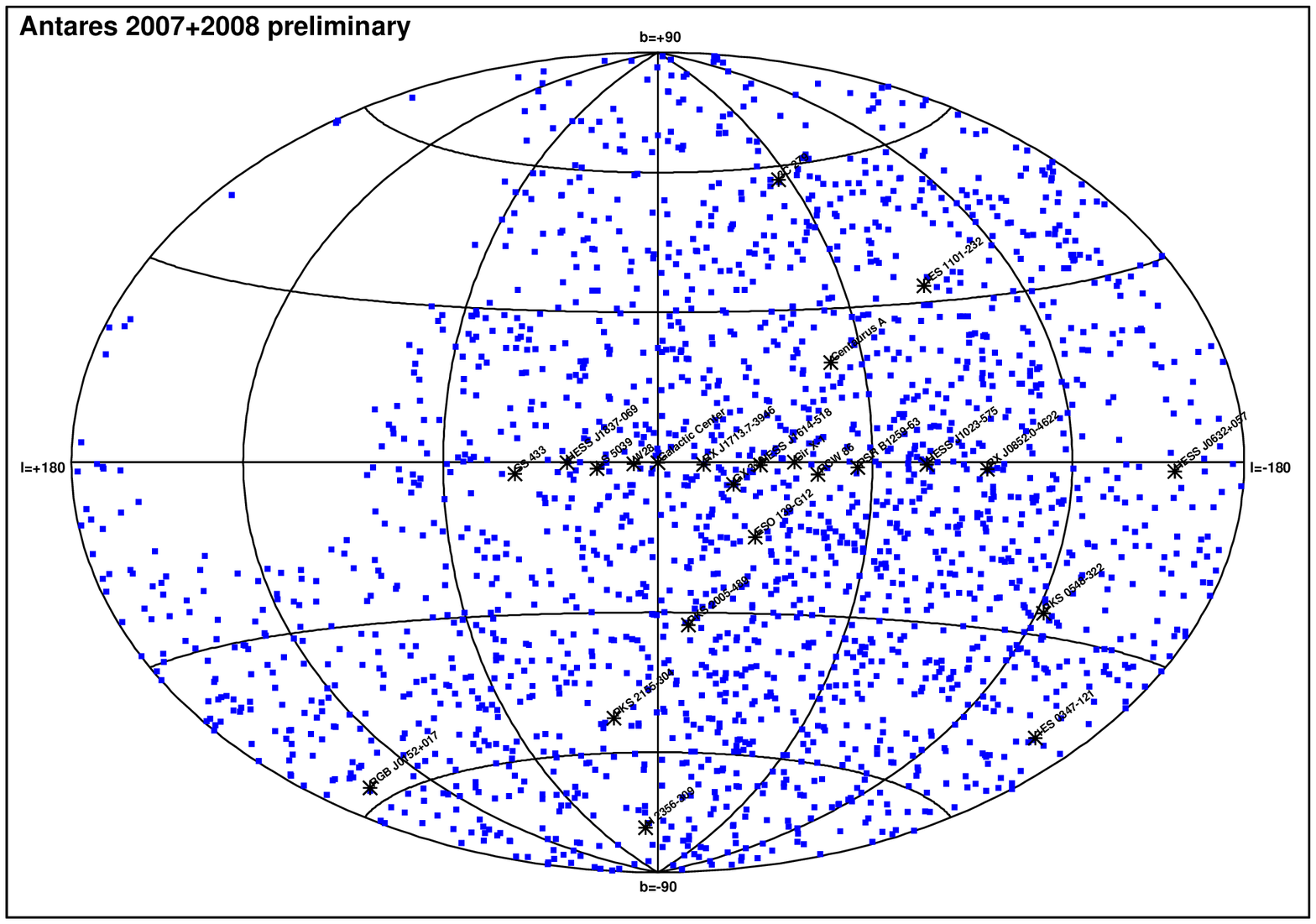}
\end{minipage}
\hfill
\begin{minipage}[t]{0.5\textwidth}
\vspace{0pt}
\includegraphics[width=\linewidth]{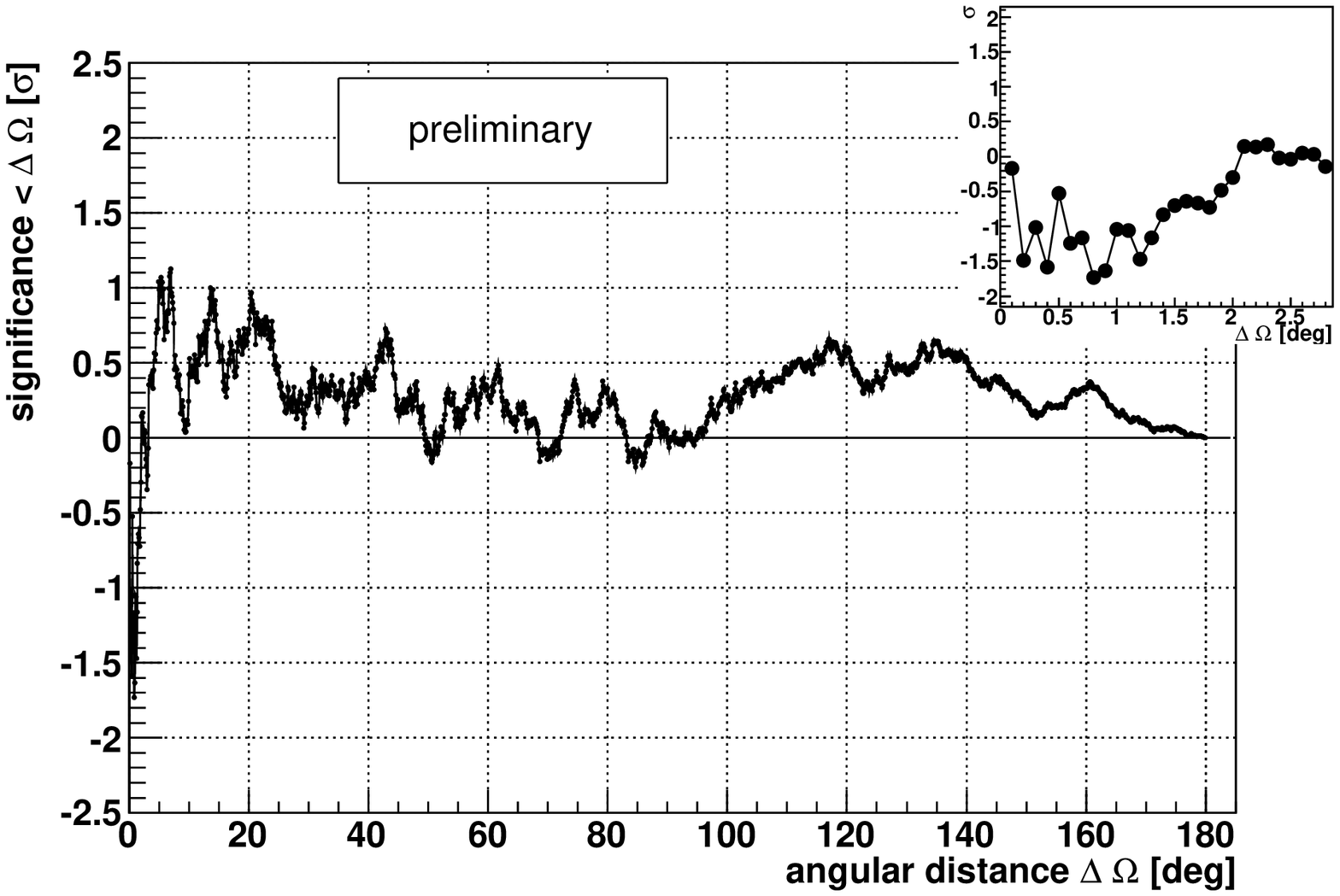}
\end{minipage}
\caption{\emph{Left}: Skymap of the 2040 neutrino candidates selected for the point source search. Stars indicate the 24 sources of the candidate 
list. \emph{Right}: Autocorrelation significance as a function of the cumulative angular scale. A maximum significance of 1.13 sigma is found for angular scales smaller than 7 degrees.}
\label{2007-8:skymap}
\end{figure}

\begin{figure}[!h]
\center
\includegraphics[width=0.55\linewidth]{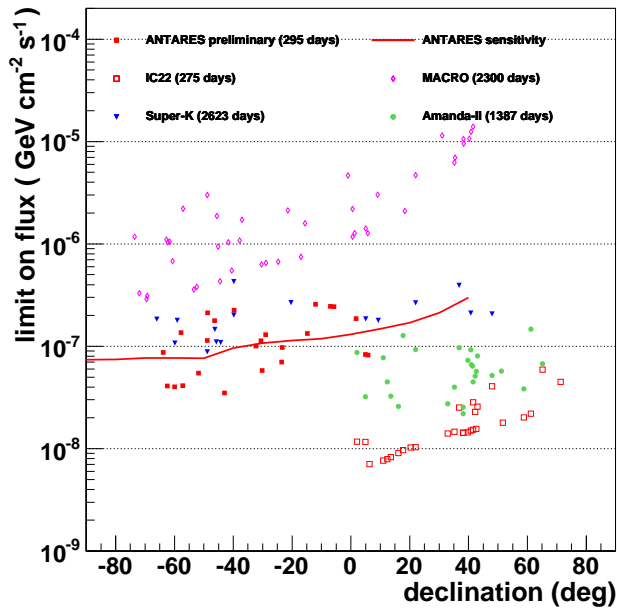}
\caption{Upper limits on the flux of HE neutrinos, assuming an E$^{-2}$ spectrum, for selected sources. 
The expected sensitivity of the ANTARES detector for a full year of data taking is also shown 
for comparison. Limits set by MACRO, AMANDA II, Super-K and IC22~\protect \cite{point_source:Refs} are superposed.}
\label{2007-8:upper_limits90}
\end{figure}

A cosmic neutrino point-like source would manifest itself as a localized excess of events on top of the 
background.\\ 
To ensure a high signal-to-noise ratio, the event reconstruction~\cite{aart} and selection are 
optimized to provide tracks with good angular resolution, in a wide energy range: the 
detector effective area is shown in figure~\ref{2007-8:plot1}-\emph{Right}.
Using data collected between 2007 and 2008, a search for point-like sources has been performed.
The integrated live-time of the data sample 
is 295 days, after data quality selection and rejection of the periods of high bioluminescence 
and high sea current. Up-going 
events, induced by muon neutrinos, are selected by imposing track quality criteria. The event selection was optimized
to achieve the best discovery potential for an assumed power-law signal with 
energy spectrum with spectral index $\gamma$ = 2.
Figure ~\ref{2007-8:skymap}-\emph{Left} shows the preliminary sample of selected events: 2040 neutrino candidates have been identified.
Simulations indicate that this sample is contaminated by a 40$\%$ of misreconstructed atmospheric muons.
Based on these events, a dedicated search for candidate sources, already known as HE gamma-rays emitters, was performed.  
This search was also completed by a full scan of the Southern sky.
Preliminary results find GX 339 as the most likely candidate source, where two events have been 
found within 1 degree of its position. The probability to 
observe this or a larger excess due to a statistical fluctuation of the background is 7$\%$.  
It can be concluded therefore that all observed excesses are compatible with the background hypothesis, 
and 90$\%$ C.L. upper limits on the neutrino flux from the considered sources have been set.
Figure~\ref{2007-8:upper_limits90} shows ANTARES upper limits as function of the 
declination, together with the expected sensitivity for 1 year 
of data taking. Results from both previous and current experiments are shown for comparison. 
The ANTARES experiment is currently providing the more stringent upper limits on the Southern sky sources, moreover these
limits are in good agreement with the expected sensitivity.\\
In a complementary analysis, the two point autocorrelation of the selected dataset has been studied. 
The applied method is independent on MonteCarlo simulations and it is sensitive to a larger variety of source morphologies. 
The reference autocorrelation distribution is determined by scrambling the data itself, so that randomized sky maps are obtained. 
The final comparison between the data and the 
reference distributions yields the significance of the differences, as a function of the cumulative angular scale, as shown in 
figure~\ref{2007-8:skymap}-\emph{Right}. The maximum significance is of 1.3 sigma, and it corresponds to angular bins smaller than 7 degrees.
\section{Search for diffuse neutrino flux}
The search for a diffuse neutrino flux, i.e. unresolved (neither in time nor in space) 
neutrino sources, is based on the search for an excess of high energy (TeV $\div$ PeV) events above 
the irreducible background of atmospheric neutrinos, whose flux is described
by a power law with a spectral index $\alpha \sim -3.5$, while several theoretical models have foreseen
an E$^{-2}$ spectrum for astrophysical neutrinos.
Atmospheric and astrophysical neutrinos can be therefore distinguished statistically on the basis of  the particle energy.
The energy estimator, called $R$, is based on hit repetitions on the PMTs, due to the different arrival 
time of  Cherenkov photons produced directly by the muon, and by delayed Cherenkov photons 
from secondary electrons and positrons dressing up the HE muon tracks. The 
average number of hit repetitions in the event is defined 
as the number of hits in the same PMT, within 500 ns 
\begin{figure}[!h]
\begin{minipage}[t]{0.5\textwidth}
\vspace{0pt}
\includegraphics[width=\linewidth]{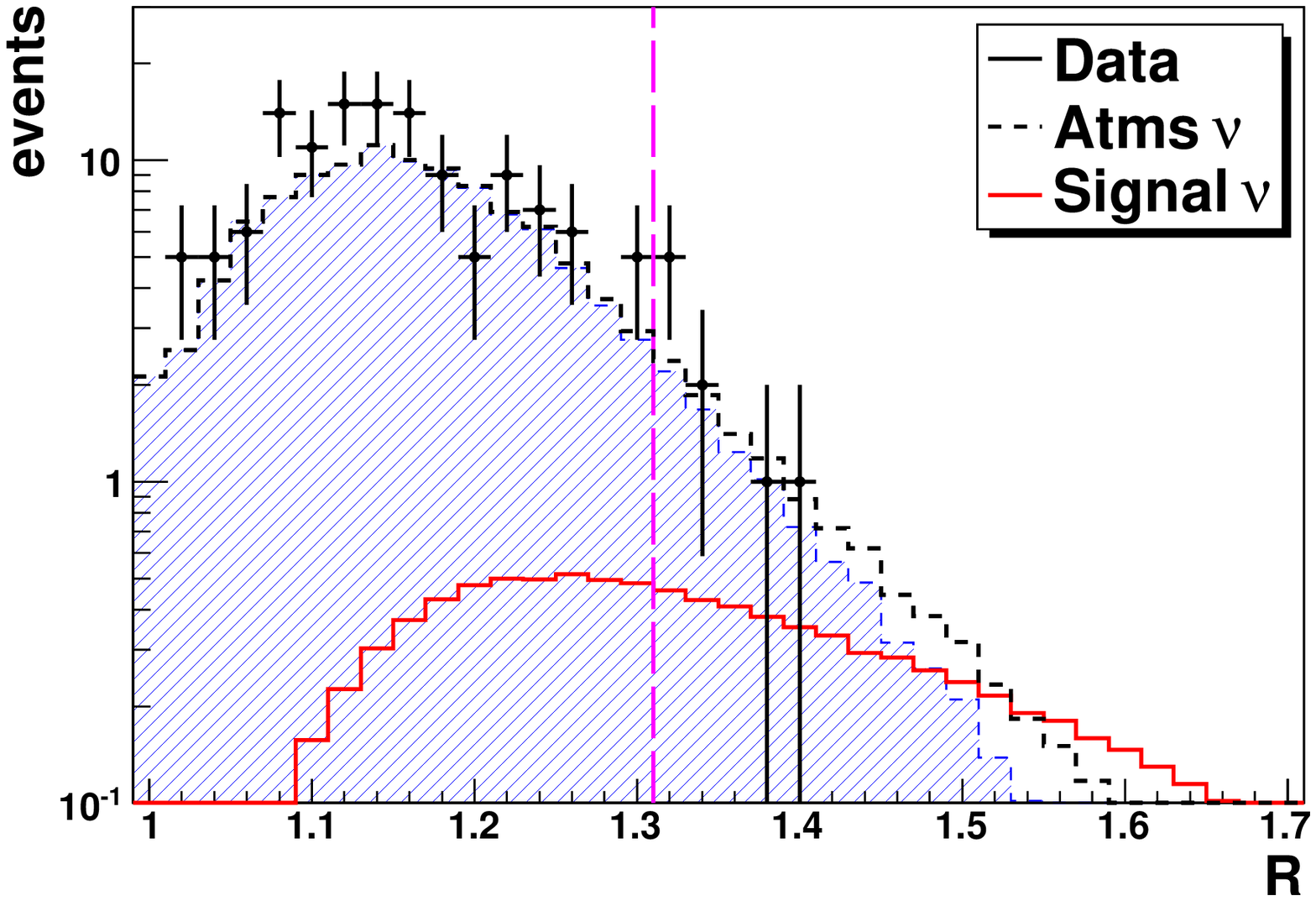}
\end{minipage}
\hfill
\begin{minipage}[t]{0.5\textwidth}
\vspace{0pt}
\includegraphics[width=\linewidth]{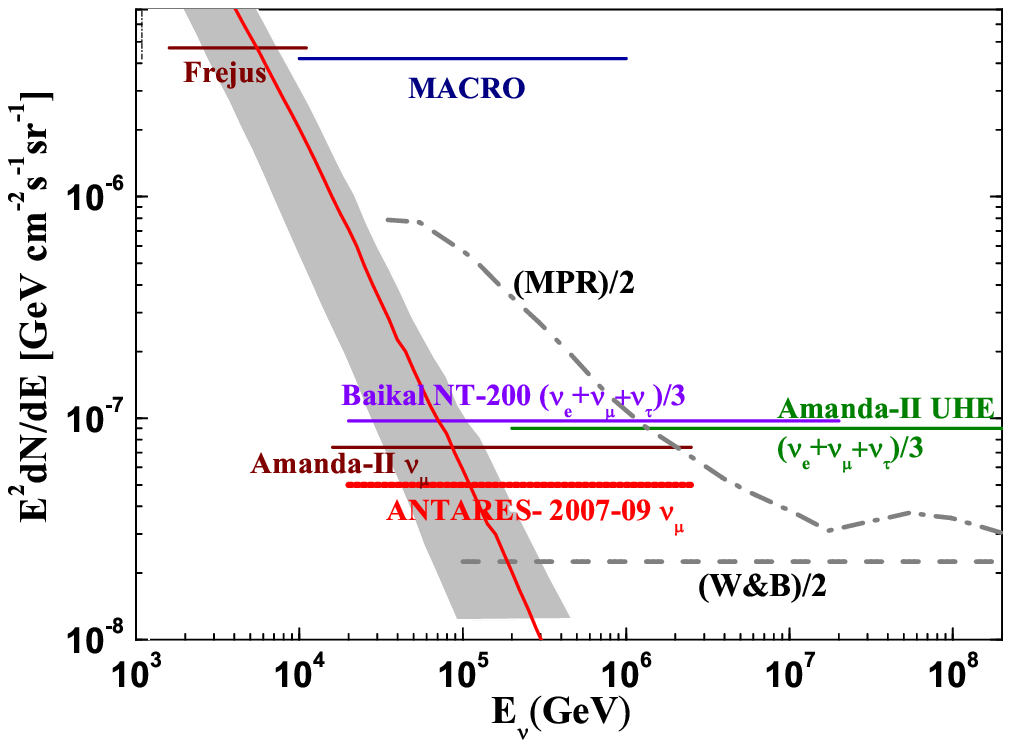}
\end{minipage}
\caption{\emph{Left}: Distribution of the energy estimator $R$ for the 134 candidate neutrino events found in the 2007-2009 
ANTARES data~\protect \cite{diffuse}, together with MonteCarlo predictions. 
Points represent data, the filled histogram and the dashed line represents simulated atmospheric neutrinos.
The signal, normalized at the upper
limit, is shown as a full line.
The optimized value $R=1.31$, that is used to discriminate between signal and background 
event, is indicated as a vertical line.
\emph{Right}: Upper limit on the diffuse neutrino flux of HE neutrinos obtained from the 2007-2009 
ANTARES data~\protect \cite{diffuse}, compared to 
theoretical predictions~\protect \cite{diffuse-theory}
and to limits set by other neutrino telescopes. See the paper for more references.}
\label{diffuse:plots}
\end{figure}    
from the earliest hit selected by the reconstruction algorithm. 
A complete analysis has been performed on data collected from December 2007 to December 2009 for a total live-time 
of 334 days~\cite{diffuse}. Figure \ref{diffuse:plots}-\emph{Left} shows the distribution of the $R$ parameter 
for the 134 candidate neutrino events found in the data sample, together with 
simulation for both background and signal neutrino events.
The number of selected events
was found to be compatible with the expected background, so that
the  90\% C.L. upper limit on the diffuse $\nu_\mu$ flux with a $E^{-2}$ spectrum is set at
$E^2\Phi_{90\%}  =   5.3 \times 10^{-8}   \  \mathrm{GeV\ cm^{-2}\ s^{-1}\ sr^{-1}} $ for 
the energy range between 20 TeV and 5 PeV, where
the energy estimator is approximately linear with log E$_{\mu}$. 
This result is shown in figure \ref{diffuse:plots}-\emph{Right}: the upper limit 
is competitive with upper limits set by other 
neutrino telescopes of comparable size and is compared to theoretical predictions~\cite{diffuse-theory}.
\section{ANTARES as an observatory for physics beyond the Standard Model}
Neutrino telescopes could also probe physics beyond the Standard Model, by detecting neutrinos 
from the annihilation of Dark Matter (DM) particles,
or exotic particles~\cite{Pavalas:2009ef} such as magnetic monopoles (MM) and slow nuclearites.\\ 
Neutrinos with energies of the order of tens of GeV could be 
produced in the annihilation of Weakly Interacting Massive Particles, e.g. 
neutralinos, which become gravitationally trapped in celestial bodies, like the Galactic Center or the Sun.
The existence of magnetic monopoles has 
been initially predicted by P. Dirac in 1931. Up-going 
\begin{figure}[!h]
\center
\includegraphics[width=0.55\linewidth]{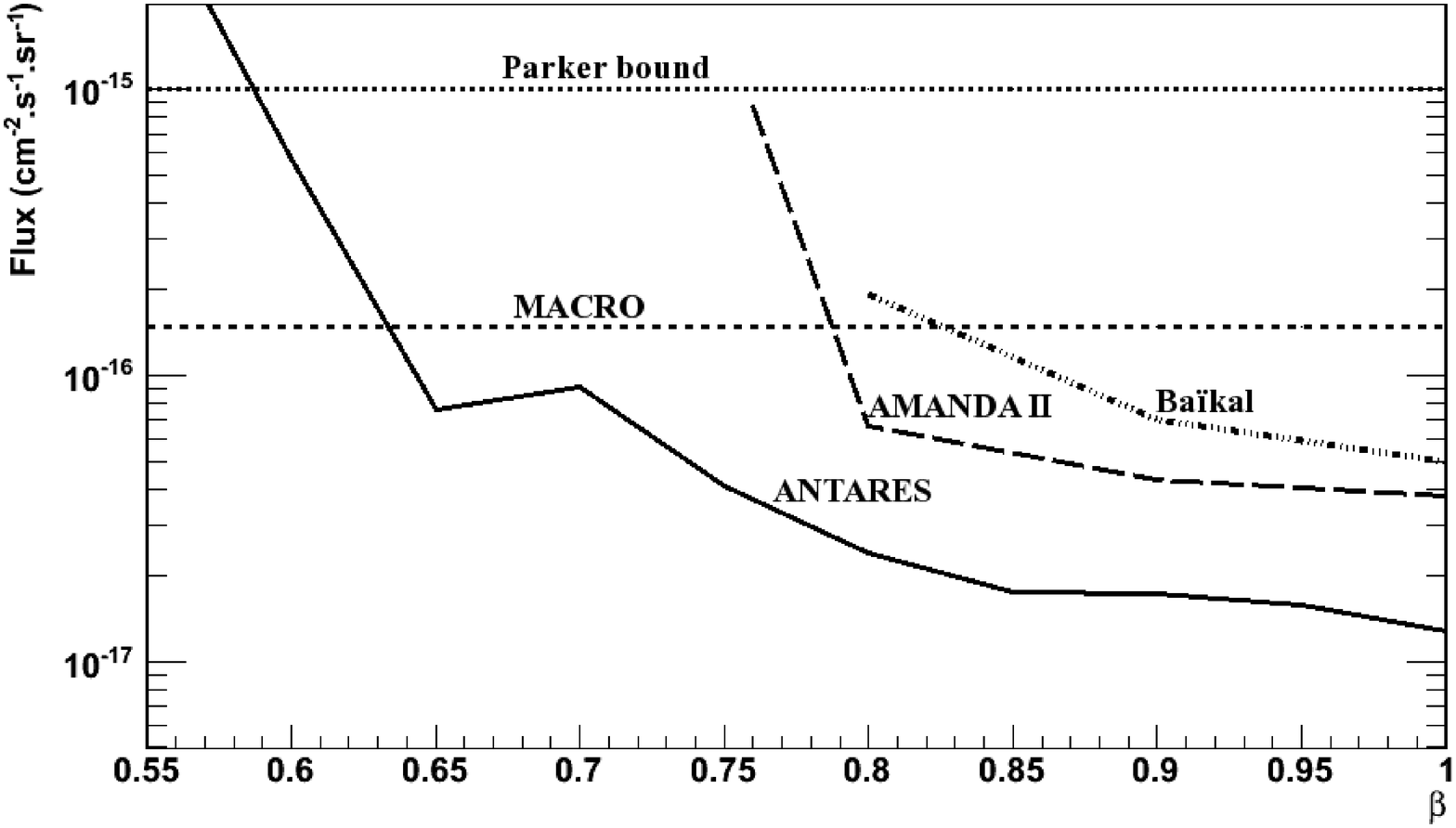}
\caption{Upper limits at 90$\%$ C.L. on the flux of fast magnetic monopoles, as 
a function of the monopoles speed: solid line indicates the preliminary result from the ANTARES 
Collaboration.}
\label{monopoles}
\end{figure}    
magnetic monopoles with masses between 10$^{10}$ and 10$^{14}$ GeV 
could be detected with 
the ANTARES detector, given their typical experimental signature: a very bright muon-like track, 
providing an amount of  photons 
that is estimated to be more than 8000 times higher than that of a muon.\\ 
The solid line in figure~\ref{monopoles} shows the preliminary limit set on the flux of MM
by the ANTARES Collaboration, for $\beta >$ 0.55. This limit is 
currently competitive with those previously established~\cite{monopoles:exp}, that are also shown for comparison.
\section{Multi-messenger approach within the ANTARES Collaboration}
The search for neutrino emission from transient sources, like for example Micro-quasars, Gamma Ray 
Bursts (GRBs)~\cite{nu-theo1} or
core collapse supernovae (ccSNe)~\cite{ando-beacom}, is 
well suited for the multi-messenger strategy. Given 
the expected small difference in the arrival time and position between photons 
and neutrinos, a very efficient rejection of the associated background can be achieved.
Due to the very low background rate, even the detection of a small number of neutrinos 
correlated with a transient source could lead to a discovery.\\
Two different detection methods have been implemented within the ANTARES Collaboration. 
The first one is the triggered search method, based on the search 
for neutrino candidates in conjunction with an accurate timing and 
positional information provided by an external source. The second 
one is the rolling search method, based on the search for high energy events or multiplets of neutrino events coming from the same 
position within a given time window,\\
GRBs are detected by gamma-ray satellites, which deliver in real 
time an alert to the Gamma-ray bursts Coordinates Network (GCN). The characteristics of this alert, mainly 
the direction and the time of the detection, are then distributed to the 
other observatories. Most gamma-ray, X-ray and optical observatories 
are capable of observing only a small fraction of the sky, for example Swift has a 1.4 sr field of view, while 
neutrino telescopes monitor essentially a full hemisphere. 
To avoid dependence on external triggers as well as to cover
a larger region of the sky, events detected with the ANTARES telescope can be used to 
trigger optical follow-up observations~\cite{kowalski-mohr}, using a Target-of-Opportunity 
(ToO) program. This method is sensitive to all transient sources producing high energy 
neutrinos. \\
The ANTARES Collaboration has developed an alert system~\cite{tatoo} that triggers the observation with a network of optical telescopes. The 
key ingredients are the use of a fast and robust reconstruction algorithm~\cite{bbfit} and the connection 
with a network of robotic telescopes with large field of view (approximately 2$^{\circ} \times$ 2$^{\circ}$), with slewing 
times of the order of tens of seconds. This is important since a GRB afterglow requires a very fast observation strategy, 
in contrary to a core collapse 
supernovae, for which the optical signal will appear several days after the neutrino signal. To be sensitive to all 
these astrophysical sources, the observational strategy is composed of a real time observation, followed by several
observations during the following month. The system is operational since 2009 and, since then, more than 30 alerts have been sent to optical
telescopes. The analysis of the optical images is under way.

\end{document}